\documentclass[aps,pra,twocolumn,groupedaddress,showpacs]{revtex4}

\usepackage{graphicx}

\begin{document}

\title{Scattering parameters for cold Li-Rb and Na-Rb collisions derived from variable phase theory}

\author{H. Ouerdane}
\author{M.J. Jamieson}
\email{mjj@dcs.gla.ac.uk}

\affiliation{Department of Computing Science, University of Glasgow, 17 Lilybank Gardens, Glasgow G12 8QQ, UK}


\begin{abstract}
We show how the scattering phase shift, the s-wave scattering length and the p-wave scattering volume can be obtained from Riccati equations derived in variable phase theory. We find general expressions that provide upper and lower bounds for the scattering length and the scattering volume. We show how, in the framework of the variable phase method, Levinson's theorem yields the number of bound states supported by a potential. We report new results from a study of the heteronuclear alkali dimers NaRb and LiRb. We consider $ab$ $initio$ molecular potentials for the X${}^1\Sigma^+$ and $a{}^3\Sigma^+$ states of both dimers and compare and discuss results obtained from experimentally based X${}^1\Sigma^+$ and $a{}^3\Sigma^+$  potentials of NaRb. We explore the mass dependence of the scattering data by considering all isotopomers and we calculate the numbers of bound states supported by the molecular potentials for each isotopomer.
\end{abstract}


\maketitle

\section{Introduction}

The study of the properties of ultra-cold trapped ensembles of atoms requires a description of very slow atomic collisions. The scattering length is a crucial quantity in the study of Bose-Einstein condensation. While there has been much interest in the behaviour of ultra-cold monatomic alkali gases, recent successes in the production of Bose-Fermi ensembles of various heteronuclear alkali dimers \cite{TRU01,SCH01,ROA02} and dual species Bose-Einstein condensates \cite{MOD02} have created further interest in studies of binary mixtures. Scattering of heteronuclear atoms at ultra-low temperature is characterised by the s-wave scattering length, $a_{\rm s}$, the effective range, $R_{{\rm eff}}$, and the p-wave scattering volume, $a_{\rm p}$, the contributions from higher angular momenta being negligible at very low energy  \cite{BLA48,BLA49,BET49,MOT65,HAD02}. There is a need for the values of the scattering lengths and scattering volumes.

In this paper we make use of the variable phase approach to potential scattering \cite{CAL67} to calculate the scattering data $a_{\rm s}$ and $a_{\rm p}$. This method is convenient and to some extent superior to the traditional calculation of phase shifts. It is advantageous compared to the usual calculations of the solutions of a second-order linear differential equation, the radial Schr\"odinger equation, that are matched to combinations of asymptotic solutions at large atomic separations, $R$. In the variable phase method one solves a first-order nonlinear differential equation, the phase equation, of the Riccati type, whose solution in the asymptotic region provides the phase shift \emph{directly}. As we show below, the phase equation is simple to manipulate in deriving properties of the scattering. 

We have shown how to use the variable phase method to compute the s-wave scattering length of a pair of colliding atoms \cite{OUE03}. We derived corrections arising from the long range interaction, accurate to at least first order in the interaction strength, and we provided upper and lower bounds to the scattering length. Use of the corrections significantly accelerates the convergence, as $R \rightarrow \infty$, to the desired scattering length, $a$, of a quantity $a(R)$ that is the scattering length {\it accumulated} at separation $R$. Here we extend our previous work to any angular momentum. Our analysis is consistent with the extension of the technique introduced for s-waves by Marinescu \cite{MAR94}. While we suggest that the the Riccati equation be solved directly, our corrections can also be applied to scattering parameters obtained from explicit computation of the wavefunction.

The variable phase method was demonstrated for s-waves \cite{OUE03} for the model potential described in reference \cite{GRI93}. Here we compute the s-wave scattering lengths and p-wave scattering volumes for the heteronuclear dimers NaRb and LiRb. We use $ab$ $initio$ molecular potentials for the two states X${}^1\Sigma^+$ and $a{}^3\Sigma^+$ \cite{KOR00}, as well as the latest X${}^1\Sigma^+$ and $a{}^3\Sigma^+$ NaRb potentials based on experimental data \cite{ZEM01,DOC04}. A difficult problem in the accurate determination of the scattering length arises from the quality of the interatomic potential. If the potential is not accurately known then it is possible that, within its error bounds, it supports a zero-energy bound state which makes the computed scattering length extremely sensitive to any change in the potential. Our results, obtained with the above potentials, are compared to other recent calculations \cite{WEI03}.

We calculate the number of bound states supported by those potentials using the simplest formulation of Levinson's theorem \cite{LEV49,CAL67}, linking the scattering phase shift, $\delta_{k,l}$, to the number of bound states, $N_{\rm b}$, {\it via} $N_{\rm b}\pi=\delta_{k,l}|_{k=0}$ where $k$ is the wavenumber of the relative motion. The scattering phase shift may be computed from the phase equation of variable phase theory with no mod[$\pi$] ambiguity.

This paper is organised as follows. In section 2 we present our mathematical model, recall results from variable phase theory \cite{CAL67,LEV63}, give detail of our derivations and show that our results may also be derived by the method of Marinescu \cite{MAR94}. In section 3 we discuss the construction of the molecular potentials \cite{KOR00,DOC04,ZEM01} and our results which we tabulate for all relevant isotopomers.

\section{Mathematical model}

\subsection{The phase equation}

The traditional method to compute the scattering phase shift, $\delta_{k,l}$, where the angular momentum in terms of the rationalised Planck's constant $\hbar$ is $l\hbar$, needs the solution of the Schr\"odinger equation at large separation. In the variable phase approach \cite{CAL67} the phase shift is obtained directly from the accumulated phase, $\delta_{k,l}(R)$ \cite{CAL67}. The accumulated s-wave phase satisfies the equation

\begin{equation} \label{eq1}
  \frac{\displaystyle {\rm d}}{\displaystyle {\rm d}R}~\delta_{k,0}(R) = -k^{-1} V(R) \sin^2\left[kR+\delta_{k,0}(R)\right],
\end{equation}

\noindent where $V(R)=2\mu {\mathcal V}(R)/\hbar^2$, ${\mathcal V(R)}$ being the interaction potential and $\mu$ the reduced mass. The phase shift is the limit of the accumulated phase as $R\rightarrow \infty$; it suffers no mod[$\pi$] ambiguity.

\subsection{Low-energy scattering}

At angular momentum $l\hbar$ the tangent of the phase $\delta_{k,l}(r)$ can be expressed in effective range theory by \cite{CAL67,LEV63,HIN71}

\begin{equation} \label{eq2}
  \tan \delta_{k,l}(R) = - k^{2l+1} \left[a_l(R) + k^2 b_l(R) + {\mathcal O}(k^4)\right],
\end{equation}

\noindent in which the coefficients $a_l(R)$ satisfy the differential equation of Riccati type

\begin{equation} \label{eq3}
\frac{\displaystyle {\rm d}}{\displaystyle {\rm d}R}~a_l(R) = \left[\alpha_l^{(+)} R^{l+1} - \alpha_l^{(-)} R^{-l}a_l(R)\right]^2 V(R)
\end{equation}

\noindent where the coefficients $\alpha_l^{(+)}$ and $\alpha_l^{(-)}$ denote $1/(2l+1)!!$ and $(2l-1)!!$ respectively where $(n+2)!!=(n+2)n!!$ with $0!!=1=1!!$. The accumulated phase, $\delta_{k,l}(R)$, is the phase shift that would be determined for a potential truncated at finite separation, $R$, and hence Eq.~(\ref{eq2}), which applies strictly for scattering by a short range potential, is valid. 

We are interested in the coefficients $a_l(R)$. For $l=0$ and $l=1$, and $R\rightarrow \infty$, these coefficients are the s-wave scattering length, $a_{\rm s}$, and the p-wave scattering volume, $a_{\rm p}$, respectively. Similar equations can be found for $b_l(R)$. In the asymptotic region the coefficient $b_0(R)$ is related to the effective range, $R_{\rm eff}$, {\it via} $b_0(\infty)=a_{\rm s}^2 R_{\rm eff}/2$. Unfortunately the equation satisfied by $b_0(R)$, which is coupled to Eq.~(\ref{eq3}) with $l=0$, is not amenable to numerical solution because of the poles that it contains; the variable phase method is not suitable for numerical calculation of the effective range but it can be used to find long range corrections to it.

\subsection{Corrections to the calculated scattering parameters}

The function $a_l(R)$ converges rather slowly with increasing separation, $R$. We need the value at infinite interatomic separation, $a_l(\infty)$. In practice we must stop the numerical calculation of the solution of Eq.~(\ref{eq3}) at a finite value, $R_{\rm c}$, of $R$. Computation time and accumulated truncation error can be large because $R_{\rm c}$ must be chosen to be sufficiently large that the long range part of the interatomic potential has neglible effect at separations greater than $R_{\rm c}$. Adopting the same analysis as in our previous work \cite{OUE03} we extend our calculations to any value of the angular momentum $l\hbar$ and show how the Riccati equation, Eq.~(\ref{eq3}), can be used to derive corrections to be applied to the calculated scattering parameters, that compensate for stopping the calculation at the finite distance $R_{\rm c}$.

We need  the contribution of the long range interaction over the range $[R_{\rm c},\infty]$. The presentation is simpler in terms of dimensionless quantities. With the substitution

\begin{equation} \label{eq4}
  {\tilde a}_l(R) = a_l(R)/B_lR_{\rm c}^{2l+1}
\end{equation}

\noindent where $B_l=\alpha_l^{(+)}/\alpha_l^{(-)}$ and the change of variable

\begin{equation} \label{eq5}
  Z=(R/R_{\rm c})^{2l+1}
\end{equation}

\noindent Eq.~(\ref{eq3}) becomes

\begin{equation} \label{eq6}
  \frac{\displaystyle {\rm d}}{\displaystyle {\rm d}R}~{\tilde a}_l = \left(Z-{\tilde a}_l\right)^2 F_l(Z),
\end{equation}

\noindent where

\begin{equation} \label{eq7}
  F_l(Z)=\frac{\displaystyle \left[R(Z)\right]^2 V(R)}{\displaystyle (2l+1)^2Z^2}.
\end{equation}

Eq.~(\ref{eq6}) has the same form as Eq.~(\ref{eq3}) of Ref.~\cite{OUE03} with $F_l(Z)$ in place of V(R). Hence, from Ref.~\cite{OUE03}, we find that for an attractive dispersion potential and sufficiently large $R_{\rm c}$, upper and effective lower bounds, ${\tilde a}_l^{(U)}$ and ${\tilde a}_l^{(L)}$ respectively, are given by

\begin{equation} \label{eq8}
  {\tilde a}_l^{({\rm U})}={\tilde a}_{l,{\rm c}}-\left(1-{\tilde a}_{l,{\rm c}}\right)^2W_{l,{\rm c}} + 2\left(1-{\tilde a}_{l,{\rm c}}\right)X_{l,{\rm c}} - 2Y_{l,{\rm c}},
\end{equation}

\noindent and

\begin{equation} \label{eq9}
  {\tilde a}_l^{({\rm L})}={\tilde a}_{l,{\rm c}} + \frac{\displaystyle -\left(1-{\tilde a}_{l,{\rm c}}\right)^2W_{l,{\rm c}} + 2\left(1-{\tilde a}_{l,{\rm c}}\right)X_{l,{\rm c}} - 2Y_{l,{\rm c}}}{\displaystyle 1+X_{l,{\rm c}}-\left(1-{\tilde a}_{l,{\rm c}}\right)W_{l,{\rm c}}},
\end{equation}

\noindent where

\begin{equation} \label{eq10}
  W_l(Z)=\int^Z F_l(Z){\rm d}Z,
\end{equation}

\begin{equation} \label{eq11}
  X_l(Z)=\int^Z W_l(Z){\rm d}Z,
\end{equation}

\noindent and

\begin{equation} \label{eq12}
  Y_l(Z)=\int^Z X_l(Z){\rm d}Z,
\end{equation}

\noindent with the subscript c denoting quantities evaluated at $R=R_{\rm c}$ or $Z=1$. Furthermore ${\tilde a}_{l,{\rm c}}$ is an upper bound.

Expressions in Eqs.~(\ref{eq7}), (\ref{eq10}), (\ref{eq11}) and (\ref{eq12}) are inconvenient for computation and, changing the variable of integration back to $R$, we find

\begin{equation} \label{eq13}
  W_l(Z)=\frac{\displaystyle 1}{\displaystyle 2l+1}~R_{\rm c}^{2l+1}\int^R R^{-2l}V(R){\rm d}R,
\end{equation}

\begin{equation} \label{eq14}
  X_l(Z)=(2l+1)~R_{\rm c}^{-2l-1}\int^R R^{2l}W_l(R){\rm d}R,
\end{equation}

\noindent and

\begin{equation} \label{eq15}
  Y_l(Z)=(2l+1)~R_{\rm c}^{-2l-1}\int^R R^{2l}X_l(R){\rm d}R.
\end{equation}

\noindent Eqs.~(\ref{eq8}), (\ref{eq9}), (\ref{eq13}) , (\ref{eq14}) and (\ref{eq15}) can be used to determine the scaled upper and lower bounds and the scaling removed, if necessary, by Eq.~(\ref{eq4}).

Equation~(\ref{eq9}) can also be determined by extending the method of Marinescu \cite{MAR94} to higher angular momenta. The radial Schr\"odinger equation is, at zero energy,

\begin{equation} \label{eq16}
  \frac{\displaystyle {\rm d}^2}{\displaystyle {\rm d}R^2}~y_l(R) - \frac{\displaystyle l(l+1)}{\displaystyle R^2}~y_l(R) - V(R)y_l(R)=0.
\end{equation}

The regular and irregular solutions of Eq.~(\ref{eq16}), to first order in the potential strength, are

\begin{equation} \label{eq17}
  u_l(R) = u_l^{(0)}(R)\left[1+B_lR_{\rm c}^{2l+1}X_l(R)\right] -2v_l^{(0)}(R)B_lR_{\rm c}^{2l+1}Y_l(R),
\end{equation}

\noindent and

\begin{equation} \label{eq18}
  v_l(R) = v_l^{(0)}(R)\left[1+B_lR_{\rm c}^{2l+1}X_l(R)\right],
\end{equation}

~\\

\noindent where $u_l^{(0)}(R)=\alpha_l^{(+)}R^{l+1}$ and $v_l^{(0)}(R)=\alpha_l^{(-)}R^{-l}$ are the regular and irregular solutions of Eq.~(\ref{eq16}) in the absence of the potential. Let $y_{l,{\rm c}}$ be the solution of Eq.~(\ref{eq16}) at $R=R_{\rm c}$. The scattering parameter, $a_{l,{\rm c}}$, is given by

\begin{equation} \label{eq19}
  a_{l,{\rm c}} = \frac{\displaystyle y_{l,{\rm c}}u_{l,{\rm c}}^{(0)}{}' - y_{l,{\rm c}}'u_{l,{\rm c}}^{(0)}}{\displaystyle y_{l,{\rm c}}v_{l,{\rm c}}^{(0)}{}' - y_{l,{\rm c}}'v_{l,{\rm c}}^{(0)}},
\end{equation}

~\\
~\\
\noindent and the scattering length, $a_l^{(1)}$, to at least first order in the potential strength is given by

\begin{equation} \label{eq20}
  a_l^{(1)} = \frac{\displaystyle y_{l,{\rm c}}u_{l,{\rm c}}' - y_{l,{\rm c}}'u_{l,{\rm c}}}{\displaystyle y_{l,{\rm c}}v_{l,{\rm c}}' - y_{l,{\rm c}}'v_{l,{\rm c}}},
\end{equation}

\noindent where the prime denotes the first derivative with respect to $R$. From Eqs.~(\ref{eq17})-(\ref{eq20}) we find that the scaled quantity $a_l^{(1)}/B_lR_{\rm c}^{2l+1}$ can be identified with $a_l^{(L)}$ of Eq.~(\ref{eq9}).

\subsection{Inverse power potential}

For the attractive inverse power potential

\begin{equation} \label{eq21}
  {\mathcal V}(R) = -C_nR^{-n},
\end{equation}

\noindent there are restrictions on $n$ and $l$; $R^{(2l+1)-(n-2)}$ must vanish as $R\rightarrow\infty$. We find the lower bound

\begin{widetext}
\begin{equation} \label{eq22}
  {\tilde a}_l^{({\rm L})} = {\tilde a}_{l,{\rm c}} - \frac{\displaystyle \frac{\displaystyle \left({\tilde \alpha}_n\right)^{n-2}}{\displaystyle 2l+1}\left[\frac{\displaystyle 1}{\displaystyle (n-2)-(2l+1)} - \frac{\displaystyle 2{\tilde a}_{l,{\rm c}}}{\displaystyle n-2} + \frac{\displaystyle {\tilde a}_{l,{\rm c}}^2}{\displaystyle (n-2)+(2l+1)}\right]}{\displaystyle 1 - \frac{\displaystyle \left({\tilde \alpha}_n\right)^{n-2}}{\displaystyle 2l+1}\left[\frac{\displaystyle 1}{\displaystyle n-2} - \frac{\displaystyle {\tilde a}_{l,{\rm c}}}{\displaystyle (n-2)+(2l+1)}\right]},
\end{equation}
\end{widetext}

\noindent where ${\tilde \alpha}_n$ is the dimensionless quantity $\left(2\mu C_n/\hbar^2\right)^{1/(n-2)}R_{\rm c}^{-1}$, and the correction that gives the upper bound, ${\tilde a}_l^{({\rm U})}$, is the numerator part of the correction in Eq.~(\ref{eq22}). The correction in the form of the numerator of Eq.~(\ref{eq22}) was derived in a different way by Szmytkowski \cite{SZM95} (whose scattering length is $B_l^{-1} \times a_l(\infty)$).

There are remaining corrections ${\tilde {\mathcal E}}^{({\rm L})}$ and ${\tilde {\mathcal E}}^{({\rm U})}$ to be made to ${\tilde a}_l^{({\rm L})}$ and ${\tilde a}_l^{({\rm U})}$. They can be derived from Eqs.~(\ref{eq20}) and ({\ref{eq24}) of Ref.~\cite{OUE03}; as noted in Ref.~\cite{OUE03} there are terms of third order in the potential strength to be added to the expression in Eq~(\ref{eq22}). The upper bound, Eq.~(\ref{eq8}), is correct to first order in the potential strength while the lower bound, Eq.~(\ref{eq9}), includes some second order terms; both need second and higher order corrections. For the potential of Eq.~(\ref{eq21}) the ratio of the upper and lower correction is given to first order asymptotically as $R\rightarrow\infty$ by

\begin{equation} \label{eq23}
  \rho = {\mathcal E}^{({\rm U})}/{\mathcal E}^{({\rm L})} = 2\left(1-\frac{\displaystyle n-2}{\displaystyle 2l+1}\right),
\end{equation}

\noindent from which we can find an improved approximation

\begin{equation} \label{eq24}
  {\tilde a}_l^{({\rm I})} = \frac{\displaystyle \rho a_l^{(L)} - a_l^{(U)}}{\displaystyle \rho - 1}.
\end{equation}

\subsection{P-wave scattering}

For $l=1$ Eq.~(\ref{eq3}) is

\begin{equation} \label{eq25}
\frac{\displaystyle {\rm d}}{\displaystyle {\rm d}R}~a_1(R) = \left[\frac{\displaystyle R^3}{\displaystyle 3} - a_1(R)\right]^2 R^{-2}V(R).
\end{equation}

\noindent whose solution, $a_1(R)$, at infinite separation is the p-wave scattering volume.

The improved approximation is, from eq.~(\ref{eq24}),

\begin{equation} \label{eq27}
  a_1^{({\rm I})} = \frac{\displaystyle 3a_1^{({\rm U})} + (2n-5)a_1^{({\rm L})}}{\displaystyle 2n-2}.
\end{equation}

S-waves are discussed in Ref.~\cite{OUE03}.

\subsection{Numerical method}

Direct finite difference methods cannot be used to solve Eq.~(\ref{eq3}) because the solution contains poles that correspond to the bound states supported by the potential. The equation can be solved by the log-derivative method \cite{MAN93,FRI94,MAN95} or by changing the variables. Substituting

\begin{equation} \label{eq28}
  a(R) = \tan [\theta(R)],
\end{equation}

\noindent and

\begin{equation} \label{eq29}
  R=\tan [\phi(R)],
\end{equation}

\noindent in Eq. (\ref{eq3}) we obtain

\begin{equation} \label{eq30}
  \frac{\displaystyle {\rm d}\theta(\phi)}{\displaystyle {\rm d}\phi} =
  \frac{\displaystyle \cos^2(\theta)}{\displaystyle \sin^2(\phi)}\left[\frac{\displaystyle \tan^3(\phi)}{\displaystyle 3} -\tan (\theta)\right]^2 V[\tan(\phi)],
\end{equation}

\noindent which we can solve over a range $[0,\phi_{\rm c}]$ by the Runge-Kutta method \cite{CLE61} where $\phi_{\rm c}$, $=\arctan R_{\rm c}$, is close to $\pi/2$.

\subsection{Advantages of the variable phase method}

We have seen that the scattering parameter $a_l$ can be evaluated by finding $a_{l,{\rm c}}$ ($=a_l(R_{\rm c})$) by propagating the solution of the first order differential equation, Eq.~(\ref{eq3}), to a suitable interatomic separation $R_{\rm c}$ and then applying the correction of Eqs.~(\ref{eq8}) and (\ref{eq9}). The variable phase method yields a simple derivation of these corrections. The corrections can also be applied to the value of $a_{l,{\rm c}}$ obtained by any other suitable method.

An advantage of the variable phase method over the traditional method is that one needs to solve a differential equation only once. In the traditional method one must solve a differential equation several times at different small wavenumbers, extract the phase shifts from the asymptotic solutions, and use the extrapolation of Eq.~(\ref{eq2}). Some experimentation is needed when choosing suitable wavenumbers.

An alternative to this method described above is to solve the zero energy Eq.~(\ref{eq16}) and extract the parameter $a_l$ from its solution by Eq.~(\ref{eq19}), as was done by Marinescu \cite{MAR94} for $a_0$. Here and in the variable phase method one must solve one differential equation once. However, because of rapid variation of the local wave number over the potential well a variable step of integration should be used to economise on computation and minimise accumulated truncation error. Adaptive methods, that select the step automatically, are well established for first order equations such as Eq.~(\ref{eq30}), derived from Eq.~(\ref{eq3}). Eq.~(\ref{eq16}) is a second order equation. Automatic step control is not so well developed for second order equations as for first order one and in using numerical methods such as that of Numerov, the step and points at which it is to be changed must be inserted explicitly. Thus the variable phase method is also advantageous over the method based on the zero energy equation.

It should be noted that the log-derivative method of solving the variable phase Eq.~(\ref{eq3}) \cite{OUE03} does not have an advantage over the Numerov method for solving the zero energy equation; the considerations concerning explicit insertion of the step apply to both and also to the traditional method described above.

\section{Numerical results for $\mbox{NaRb}$ and $\mbox{LiRb}$}

\subsection{The molecular potentials}

We compute scattering parameters for LiRb and NaRb. We use various potentials and compare and discuss results. Korek {\it et al.} \cite{KOR00} calculated $ab$ $initio$ molecular potentials for the X${}^1\Sigma^+$ and $a{}^3\Sigma^+$ states of LiRb and NaRb. Docenko {\it et al.} \cite{DOC04} constructed an X${}^1\Sigma^+$ NaRb potential from experimental data and from the $a{}^3\Sigma^+$ state potential made by Zemke {\it et al.} \cite{ZEM01}. In the long range dispersion potentials we used the van der Waals coefficients computed by Derevianko {\it et al.} \cite{DER01,POR03} with $C_6$ from Ref.~\cite{DER01} and $C_8$ and $C_{10}$ from Ref.~\cite{POR03}.

There is difficulty in matching an analytic long range potential to the short range part which is usually available only in tabulated form. Any artificial discontinuity at the matching point can greatly affect the calculated scattering data. For LiRb the matching point that gives the smoothest fit is at a separation of 40 bohr. We found that within the precision of the data of Korek {\it et al.} \cite{KOR00} the singlet and triplet potentials do not differ beyond 30 bohr and hence we neglected exchange beyond the matching point at 40 bohr, which is outside the typical LeRoy radius \cite{LER73} of 10-15 \AA~ of alkali dimers. For NaRb we used the short range potentials of Korek {\it et al.} \cite{KOR00} and matched at 30 bohr to the long range dispersion form evaluated with the dispersion coefficients of Derevianko {\it et al.} \cite{DER01,POR03}.

The computed scattering parameters crucially depend on the potential. It is important to compare results obtained with various molecular potentials. The most recent short range LiRb potential is that of Korek {\it et al.} \cite{KOR00}. However Docenko {\it et al.} \cite{DOC04} provided new measurements pertinent to the X${}^1\Sigma^+$ molecular state of the NaRb molecule and Zemke {\it et al.} \cite{ZEM01} also made calculations for the $a{}^3\Sigma^+$ state of the NaRb molecule. We use calculations with these potentials in a comparison with the recent results of Weiss {\it et al.} \cite{WEI03}.

We include exchange explicitly only for the $a{}^3\Sigma^+$ state of the NaRb molecule for which the point for matching to the long range potential is within the LeRoy radius. We represent exchange by the expression of Smirnov and Chibisov \cite{SMI65} in the formulation given by Weiss \emph{et al} \cite{WEI03}. 

\subsection{Numerical results}

\begin{figure}
{\scalebox {0.31}{\includegraphics{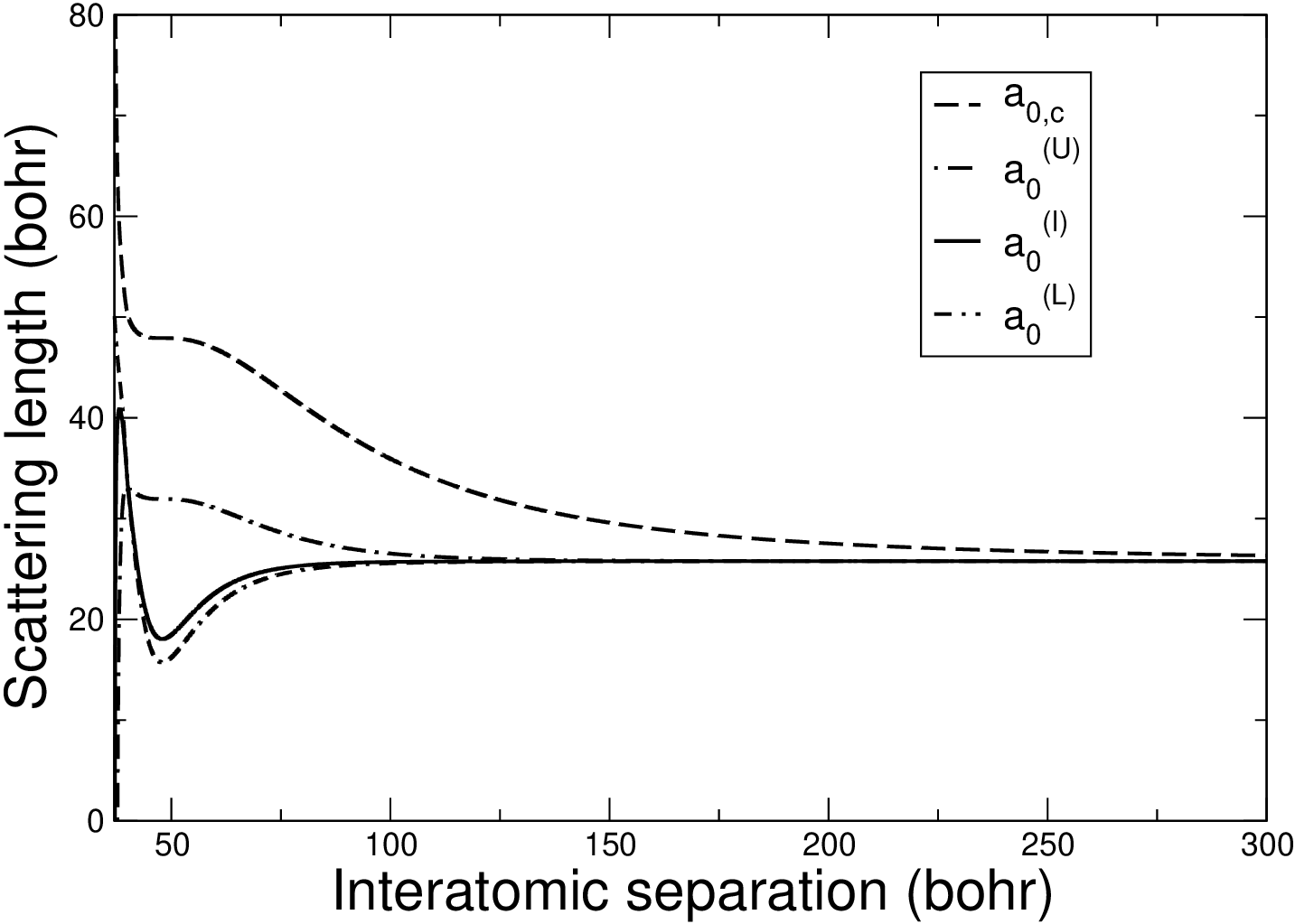}}}
\caption{\label{fig:eps1} Accumulated and corrected scattering lengths for ${}^6\mbox{Li}{}^{85}\mbox{Rb}$ in the triplet state $\left[a{}^3\Sigma^+\right]$, as functions of the interatomic distance $R$.}
\end{figure}

We tabulate the calculated scattering length data, $a_{\rm s}$ and $a_{\rm p}$, and the numbers of bound states, $N_{\rm b}$, supported by the molecular potentials. The values are those obtained from the analyses above including corrections for $a_{\rm p}$, those that we obtained from our previous analyses  for $a_{\rm s}$ \cite{OUE03} and those obtained from Levinson's theorem applied to the solution of Eq.~(\ref{eq1}) for $N_{\rm b}$. We compare our results with those for NaRb of Weiss {\it et al.} \cite{WEI03}.

\subsubsection{Study of LiRb}

We  used the following values for the masses of the isotopes, all in atomic mass, units obtained from the National Institute of Standards and Technology (NIST): $M_{\rm {{}^6Li}}=6.0151223$, $M_{\rm {{}^7Li}}=7.016004$, $M_{\rm {{}^{85}Rb}}=84.9117893$ and $M_{\rm {{}^{87}Rb}}=86.9091835$. For the singlet state X${}^1\Sigma^+$ we see that the nature of the isotopomer influences the value of the scattering length and its sign; the scattering length is positive for the smaller mass and negative for the bigger mass. However a change in the Rb mass changes the value of the scattering length but not its sign. The order of magnitude of the various values of the scattering length does not change. The greater influence of mass changes of Li compared to Rb is caused by the greater sensitivity of the reduced mass of the dimer to changes in the mass of the lighter atom. The scattering volume is more stable with its sign remaining unchanged and its values having the same order of magnitude and is less sensitive to a change of the masses of the atoms. The number of bound states is  sensitive to the Li mass.\\

\begin{figure}
{\scalebox {0.31}{\includegraphics{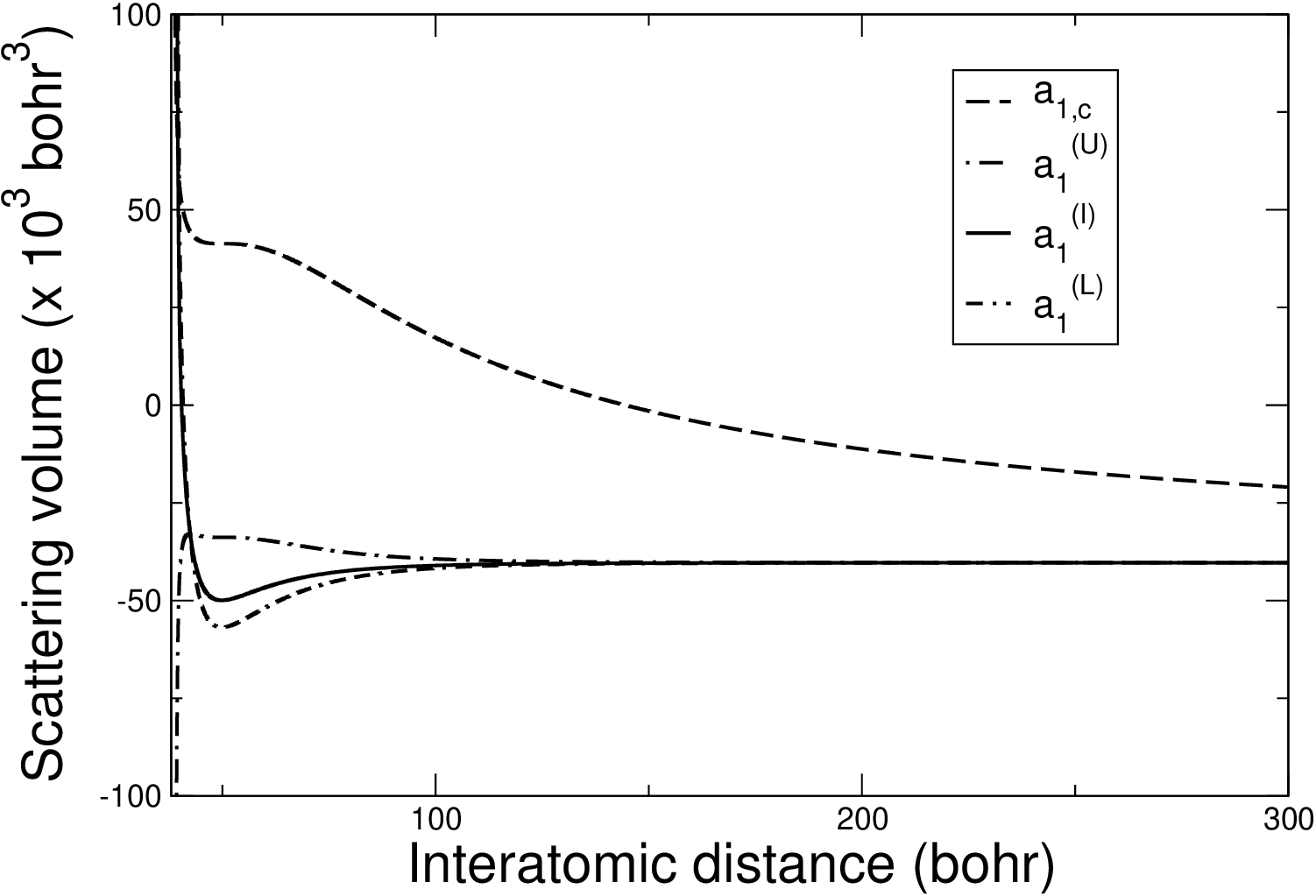}}}
\caption{\label{fig:eps2} Accumulated and corrected scattering volumes for ${}^6\mbox{Li}{}^{85}\mbox{Rb}$ in the triplet state $\left[a{}^3\Sigma^+\right]$, as functions of the interatomic distance $R$.}
\end{figure}

The $a{}^3\Sigma^+$ potential well is much shallower than the X${}^1\Sigma^+$ well. Therefore the value of the reduced mass of the LiRb molecule does not dramatically affect the scattering length (although it becomes slightly negative for the heaviest isotopomer). The scattering volume keeps its sign and there is a change of only one bound state when the LiRb dimer is composed of either of the Rb isotopes and the heavier of the Li isotopes. However there is a qualitatively opposite behaviour in the changes of $a_{\rm p}$ when the mass of Li is greater as compared with the results obtained for the X${}^1\Sigma^+$ state; the scattering volume is greater in magnitude for the smaller values of the mass of Li in the X${}^1\Sigma^+$ state and smaller for the greater value of the mass of Li in the $a{}^3\Sigma^+$ state.

The semiclassical analysis of Gribakin and Flambaum ~\cite{GRI93} can be invoked to explain the variation of scattering length with mass. The semiclassical scattering length depends only on the action integral which, for a given potential, is proportional to the square root of the reduced mass. The semiclassical formulas for the scattering length and the number of bound states, combined with the dispersion parameters of the potential and the computed scattering length, allow us to find the action integral appropriate to our quantum calculations. The action integral divided by the square root of the reduced mass thus calculated should be independent of isotopomer. This quantity obtained from our quantal results varies by less than 2\% over the range of LiRb isotopomers considered, which is consistent with semiclassical theory. It has the values 1.6 and 0.49 atomic units for the X$^1\Sigma^+$ and the a$^3\Sigma^+$ potentials respectively.

Figures 1 and 2 illustrate the convergence of the approximations to the scattering parameters. Figure 3 shows the accumulated phase shift, demonstrating how it is related to the number of bound states.

New $ab$ $initio$ or spectroscopically determined LiRb molecular potentials for the X${}^1\Sigma^+$ and $a{}^3\Sigma^+$ states will be extremely useful to allow comparison of the results displayed in Table~\ref{tab:table3} and hence to confirm the physical implications of the signs of the scattering parameters, and the numbers of bound states. This is important for NaRb; when we compare values of $a_{\rm s}$, $a_{\rm p}$ and $N_{\rm b}$ computed from three different potentials for each molecular state we find discrepancies.

\begin{figure}
{\scalebox {0.31}{\includegraphics{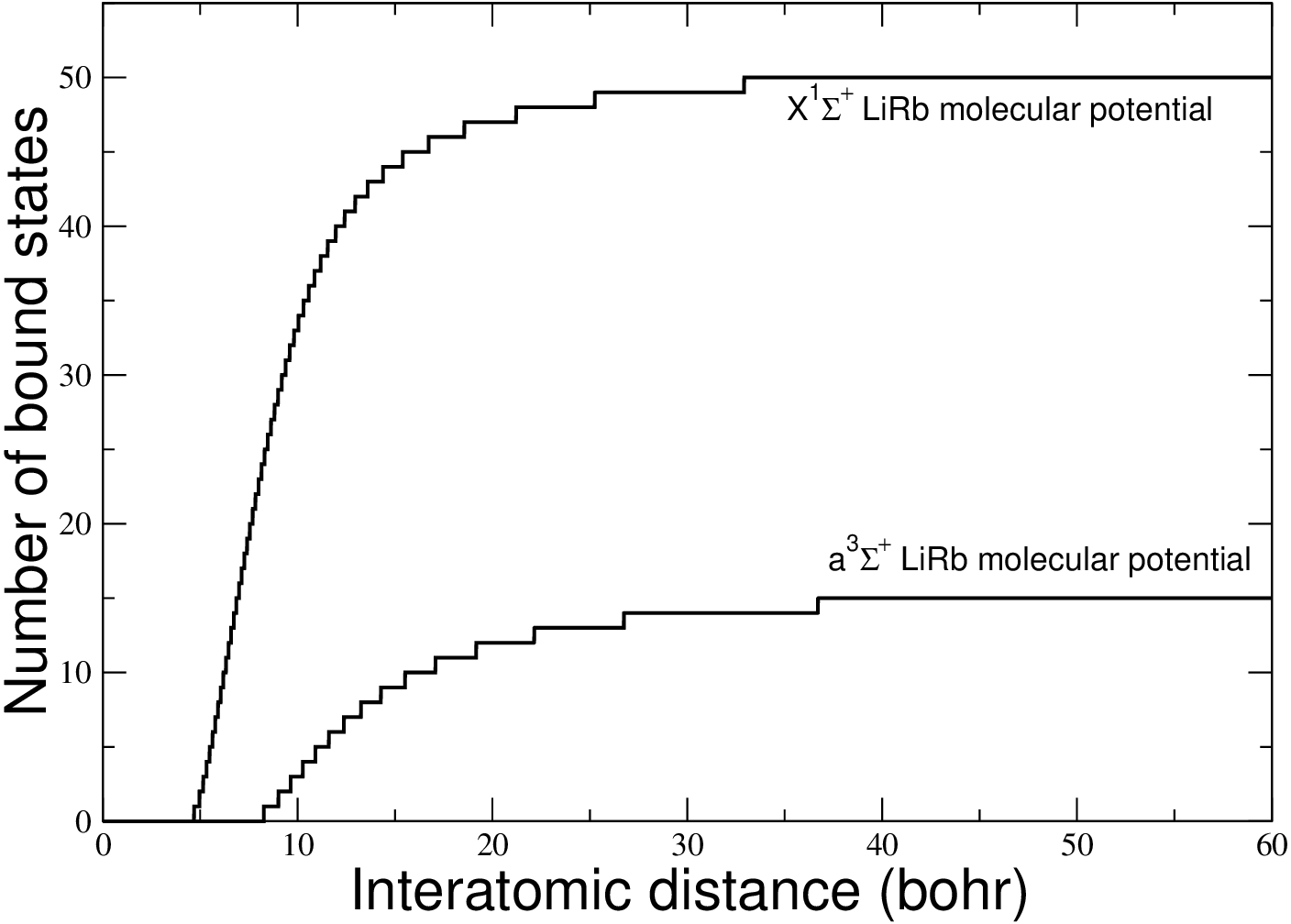}}}
\caption{\label{fig:eps3} Accumulated phase shifts (in units of $\pi$ radians) as functions of the interatomic distance $R$ and number of bound states for ${}^6\mbox{Li}{}^{85}\mbox{Rb}$ in the singlet and triplet states.}
\end{figure}

\begin{table}
\caption{\label{tab:table3}The scattering length (in bohr), scattering volume (in bohr${}^3$) and number of bound states for LiRb.}
\begin{ruledtabular}
\begin{tabular}{cccc}
LiRb $\left[X{}^1\Sigma^+\right]$ &$a_{\rm s}$ &$a_{\rm p}$ &$N_{\rm b}$\\
\hline
${}^6\mbox{Li}{}^{85}\mbox{Rb}$ & -40 & -97720 & 50\\
${}^6\mbox{Li}{}^{87}\mbox{Rb}$ & -64 & -105899 & 50\\
${}^7\mbox{Li}{}^{85}\mbox{Rb}$ & 26 & -44698 & 54\\
${}^7\mbox{Li}{}^{87}\mbox{Rb}$ & 18 & -59648 & 54\\
\hline\hline
LiRb $\left[a{}^3\Sigma^+\right]$ &$a_{\rm s}$ &$a_{\rm p}$ &$N_{\rm b}$\\
\hline
${}^6\mbox{Li}{}^{85}\mbox{Rb}$ & 26 & -40309 & 15\\
${}^6\mbox{Li}{}^{87}\mbox{Rb}$ & 24 & -43583 & 15\\
${}^7\mbox{Li}{}^{85}\mbox{Rb}$ & 3 & -78665 & 16\\
${}^7\mbox{Li}{}^{87}\mbox{Rb}$ & -0.25 & -82200 & 16\\\end{tabular}
\end{ruledtabular}
\end{table}

\subsubsection{Study of NaRb}

We took the value for the mass of ${}^{23}\mbox{Na}$ given by NIST as 22.98976967 atomic mass units. There are two isotopomers.

Our results, displayed in Table~\ref{tab:table4}, computed with the X${}^1\Sigma^+$ potential of Docenko {\it et al.} \cite{DOC04} are in excellent agreement with those of Weiss {\it et al.} \cite{WEI03} for $^{23}$Na$^{85}$Rb. Weiss {\it et al.} \cite{WEI03} claimed that for NaRb the $ab$ $initio$ potentials of Korek {\it et al.} \cite{KOR00} may not be appropriate for the calculation of scattering lengths as they do not agree with the potentials determined from spectroscopic measurments.

The values of $a_{\rm s}$ and $a_{\rm p}$ depend again on the mass of the molecule. However the number of bound states is not influenced by the choice of  NaRb isotopomer because the Rb isotopes differ little in mass. The sign of the scattering length is not influenced by the choice of  NaRb isotopomer for the singlet potential but, for the triplet state, there is a qualitative difference  between the results computed from the potential of Korek {\it et al.} \cite{KOR00} (negative scattering length) and those obtained from the potentials of Weiss {\it et al.} \cite{WEI03} and of Zemke {\it et al.} \cite{ZEM01} (positive scattering lengths). The numbers of bound states computed from potentials in Refs.~\cite{DOC04,ZEM01} and \cite{WEI03} agree whereas we find fewer bound states supported by both triplet and singlet potentials of Ref.~\cite{KOR00}. This explains the change of sign of the scattering length.

\begin{table}
\caption{\label{tab:table4}The scattering length (in bohr), and number of bound states for NaRb.}
\begin{ruledtabular}
\begin{tabular}{ccccccccc}
NaRb $\left[X{}^1\Sigma^+\right]$ &$a_{\rm s}$\footnotemark[1] &$a_{\rm s}$\footnotemark[2] &$a_{\rm s}$\footnotemark[3] &$a_{\rm s}$\footnotemark[4] &$N_{\rm b}$\footnotemark[1] &$N_{\rm b}$\footnotemark[2] &$N_{\rm b}$\footnotemark[3]&$N_{\rm b}$\footnotemark[4]\\
\hline
${}^{23}\mbox{Na}{}^{85}\mbox{Rb}$ & 174 & 62 & 167 & 178 & 83 & 76 & 83 & 82\\
${}^{23}\mbox{Na}{}^{87}\mbox{Rb}$ & 84 & 29 & 55 & 87 & 83 & 76 & 83 & 82\\
\hline\hline
NaRb $\left[a{}^3\Sigma^+\right]$ &$a_{\rm s}$\footnotemark[1] &$a_{\rm s}$\footnotemark[2] &$a_{\rm s}$\footnotemark[3] &$a_{\rm s}$\footnotemark[4] & $N_{\rm b}$\footnotemark[1] & $N_{\rm b}$\footnotemark[2] &$N_{\rm b}$\footnotemark[3] &$N_{\rm b}$\footnotemark[4]\\
\hline
${}^{23}\mbox{Na}{}^{85}\mbox{Rb}$ & - & - 51 & 59 & 105 & - & 18 & - & 22\\
${}^{23}\mbox{Na}{}^{87}\mbox{Rb}$ & - & -102 & 51 & 91 & - & 18 & - & 22\\\end{tabular}
\end{ruledtabular}
\footnotetext[1]{Calculated from potential in Ref.~\cite{DOC04}.}
\footnotetext[2]{Calculated from potential in Ref.~\cite{KOR00}.}
\footnotetext[3]{Results from Ref.~\cite{WEI03} or other reference therein.}
\footnotetext[4]{Calculated from potential in Ref.~\cite{ZEM01}.}
\end{table}

We compare the scattering volumes computed from the potentials of Korek {\it et al.} \cite{KOR00} with those computed from the potentials of Zemke {\it et al.} \cite{ZEM01}. A striking difference is the opposite signs for both singlet potentials. In a semiclassical analysis similar to that described above for LiRb we found that the action integral divided by the square root of the reduced mass obtained from our quantal results varies by less than $\frac{1}{3}$\% for the NaRb isotopomers considered. In atomic units it has the values 1.4 and 0.32 for the X$^1\Sigma^+$ and the a$^3\Sigma^+$ potentials of Korek {\it et al.} \cite{KOR00} respectively, the values 1.5 and 0.39 for the X$^1\Sigma^+$ and the a$^3\Sigma^+$ potentials of Zemke {\it et al.} \cite{ZEM01} respectively and the value 1.5 for the X$^1\Sigma^+$ potential of Docenko {\it et al.} \cite{DOC04}.

\begin{table}
\caption{\label{tab:table5}The scattering volume (in bohr${}^3$) of NaRb.}
\begin{ruledtabular}
\begin{tabular}{ccccc}
NaRb $\left[X{}^1\Sigma^+\right]$ &$a_{\rm p}$\footnotemark[1] &$a_{\rm p}$\footnotemark[2] &$a_{\rm p}$\footnotemark[3]&$a_{\rm p}$\footnotemark[4]\\
\hline
${}^{23}\mbox{Na}{}^{85}\mbox{Rb}$ & -681980 & 52815 & - & -663097\\
${}^{23}\mbox{Na}{}^{87}\mbox{Rb}$ & 373425 & -114509 & - & 459310\\
\hline\hline
NaRb $\left[a{}^3\Sigma^+\right]$ &$a_{\rm p}$\footnotemark[1] &$a_{\rm p}$\footnotemark[2] &$a_{\rm p}$\footnotemark[3]&$a_{\rm p}$\footnotemark[4]\\
\hline
${}^{23}\mbox{Na}{}^{85}\mbox{Rb}$ & - & -235850 & - & 3085301\\
${}^{23}\mbox{Na}{}^{87}\mbox{Rb}$ & - & -266190 & - & 615416\\
\end{tabular}
\end{ruledtabular}
\footnotetext[1]{Calculated from potential in Ref.~\cite{DOC04}.}
\footnotetext[2]{Calculated from potential in Ref.~\cite{KOR00}.}
\footnotetext[3]{Results from Ref.~\cite{WEI03} or other reference therein.}
\footnotetext[4]{Calculated from potential in Ref.~\cite{ZEM01}.}
\end{table}

\section{Conclusion}

We have shown that the variable phase method is appropriate for the study of very slow atomic collisions as it allows \emph{direct} calculation of scattering parameters and the number of bound states. The method is convenient. It let us derive corrections that form upper and lower bounds for $a_{\rm s}$ and $a_{\rm p}$. The corrections are simple and we suggest that use of the first-order correction with a moderate value of $R_c$ is sufficient to evaluate $a_{\rm s}$ and $a_{\rm p}$. We demonstrated the simplicity and efficiency of the phase method with our study of Li-Rb and Na-Rb collisions, and we reported new scattering data for these systems.

\begin{acknowledgments}

We are pleased to thank Dr. D. Vrinceanu for discussions about the variable phase method, and the referee for some useful comments. This work was supported by the Engineering and Physical Sciences Research Council, UK.

\end{acknowledgments}

\bibliography{pra}

\end{document}